# Compression and Quantitative Analysis of Buffer Map Message in P2P Streaming System

C.X., Li, C.J. Chen, D.M Chiu

Abstract
   BM compression is a straightforward and operable way to reduce buffer message length as well as to improve system performance. In this paper, we thoroughly discuss the principles and protocol progress of different compression schemes, and for the first time present an original compression scheme which can nearly remove all redundant information from buffer message. Theoretical limit of compression rates are deduced in the theory of information. Through the analysis of information content and simulation with our measured BM trace of UUSee, the validity and superiority of our compression scheme are validated in term of compression ratio.

I Background
   In recent years, P2P streaming media system has gotten explosive development. As an important and hot section in P2P world, the P2P streaming system is attracting a lot research eyes. Many works such as [4]-[13] try to characterize the system's features in nature based on network measurement. With deeper understanding the system, some system models are proposed to improve the system performance in terms of information broadcast speed, network sharing efficiency and startup delay, etc. Although some of the models, which are based on real P2P streaming applications, are engineering feasible, majority of them stay on the authors' desks due to too far away to the reality.
   In 2008, Libo [1] found that increasing the exchange rate of peers' buffer map message (BM) could help chunk diffusion. This find opens another convenient door for us to improve the P2P streaming system's performance. Naturally, decreasing the cycle of BM exchange will increase the BM protocol overhead in term of network traffic, which will lead to uncertain negative effects on system performance. Nearly at the same time, we found that some mainstream ASPs including PPLive and UUSee have reduced their BM length while kept buffer width and BM exchange cycle unchanged. Our findings confirms that even under the same BM exchange cycle as before, the original BM of more than 80 bytes length has brought non-ignorable overhead. It gives us the BM overhead baseline, which any performance optimization solution must be kept below. Under this principle, the original BM must be scaled down.
   In general, there are several ways to shorten the BM, such as decreasing the buffer width, enlarge the chunk size, and compressing the bitmap. Both buffer width and chunk size are the system design parameters, and any slight adjust on them may affect the whole situation and bring systematic challenges. In practice, software engineers always try to do as best as they can, to carefully and empirically tune the whole system into the best situation with the most suitable parameters. On the other hand, following a different way from the adjustment of the buffer width and chunk size, BM compression can further improve the improvements in the adjustments. Furthermore, no any design and performance risks will be involved in BM compression.
   Data compression is not a new topic, but the existing algorithms provide the universal methods which are not specifically designed for P2P applications. Of course the BM can be compressed with traditional methods, but whether it is well compressed? Based on full consideration to the unexplored feathers of BM, we try to design more powerful compression scheme, to theoretically discuss the difference of all the BM compression schemes, and to design the optimal BM compression approach. All such questions are very interesting and important not only to us but also (especially) to the system's designer and developer. To the best of our knowledge, the BM compression issues have never been thought about and discussed seriously and thoroughly.

In the paper, we discuss both how to compression BM message and how well a best compression method can be reached. Firstly, after a brief introduction of known schemes, including the single BM scheme (SBMS) used in practice and the single peer's BM scheme (SPBMS) proposed in [2], we present a more efficient compression method namely paired peer's BM scheme (PPBMS) which can thoroughly remove the redundant information. Secondly, achievable compression rate for each discussed methods are estimated under the stationary buffer filling assumption. We assume the buffer filling is a stationary process following certain s curve, and deduce different scheme's compression rate. According to our analysis result with UUSee S curve, PPBMS has a significant advantage over other schemes. At last, different achievable compression schemes are simulated with our actual measured UUSee BM trace and the quantitative results validate the validity and superiority of our PPBMS.

In the paper, we discuss different kind of BM compression schemes including the single BM scheme (SBMS) used in practice, the single peer's BM scheme (SPBMS) proposed in [2], and present a more efficient compression method namely paired peer's BM scheme (PPBMS) which can thoroughly remove the redundant information. Based on our previous research on the buffer filling behavior, as well as the BM exchange feature, the limit for each scheme, which is one of the key issues seriously considered by system and protocol designer, is deduced for the first time from the perspective of the information theory. Base on the limit, we can design a better compression protocol to improve the P2P system performance. According to our analysis, PPBMS limit has a significant advantage over other schemes. At last, different BM schemes are simulated with our actual measured UUSee BM trace and the quantitative results validate the validity and superiority of our PPBMS.

The remaining sections are arranged as follows: in section ii, the BM structure and its exchange process are depicted, and existing BM compression method are explained; our PPBMS are present in section iii; in section iv we present a analysis model to deduce the limit of different BM compression solution, and as a sample, the limits and overhead of all schemes in UUSee are calculated and compared; in section v different coding schemes are simulated and compared with our measured BM trace of UUSee; at last, section vi is a summary.

## II Overview of BM and its current situation

A P2P live streaming system uses few servers to support large number of audiences (named as peer). In current typical P2P streaming media systems, streaming content is cut into continuous blocks (or chunks) transmitted between peers, each of which is marked by a unique sequence number called chunk id in many paper. Content server (or seeder) of the system injects chunks (required by peers) one by one into the system and each peer caches the chunks in a buffer organized in chunks. A buffer message (BM) is introduced to abstractly descript this buffer status. Typically, BM consists of an *offset*, which is the ID of the chunk at the buffer head, and a sequence of {0,1}, in other words, a bitmap of the buffer to depict the download scenario in this buffer. A value of 1(0) at the $i^{th}$ position indicates that the chunk with an $ID_{offset+i-1}$ has been (has not been) buffered in the buffer. The length of bitmap is the peer's buffer width. Program content's sharing and exchanging between peers heavily rely on BM reported by their counterparts.

BM protocol overhead is no longer a negligible part in P2P streaming system. The BM protocol traffic comes from the BM size and its exchange frequency. As we know, the ability to resist network stirring becomes stronger as the buffer width becomes larger. Thus, in order to improve the capability of continuous playback and give user better watching experience, large buffer solution is usually adopted in current P2P streaming system. On the other side of the coin, large buffer leads to high playback delay and big BM size. Usually the former isn't a fatal performance factor because the audiences are insensitive to the delay in one-way broadcasting video program. However, the large BM increases the protocol overhead. Libo in his paper [1] finds the relation between the cycle of BM exchange and the chunk diffusion speed. For better network sharing environment, peer wishes faster BM updating while it will leads to more overhead.

What is the right size and right cycle is a very complex issue. In engineering practice, engineers continue to adjust it to a balance. According to our measurement, for UUSee and PPLive respectively, the BM size is about 80 bytes and 250 bytes in their early version, and the exchange cycle is about 5s and 4s. Assuming a peer keeps 30 connections with other peers at the same time on average, the protocol overhead is about 7.68kb/s and 30kb/s respectively. Considering the general case of 512kb/s ADSL access speed with 400kb/s video stream, such an overhead cannot be overlooked. In 2009 we find the mainstream ASPs (PPLive and UUSee) have reduced their BM size while kept buffer width and BM exchange cycle unchanged. Our find confirms that even under the same BM exchange cycle as before, the original BM of more than 80 bytes length has brought non-ignorable overhead. It also gives us the baseline of BM traffic overhead. The BM traffic cost in any performance optimization solution must not be larger than the baseline.

BM compression is a straightforward and operable way to reduce this overhead as well as to improve system performance. However, to the best of our knowledge, this problem has never been thought about and discussed seriously and thoroughly. Following we will discuss some key and interesting issues, including different BM compression methods，the theoretic compression ratio and the optimal achievable coding scheme design.

III BM compression solutions

In early P2P streaming system, the original BM describing the content buffer is exchanged between peers without any compression. In further engineering practice and study, people find such an original BM scheme leads to heavy overhead and try to look for compression solutions. As described above, BM contains two most important elements: offset and bitmap. Offset usually is 4-byte in size, and bitmap includes many bytes. In our following discussion, we mainly focus on the compression of *bitmap* and assume *offset* is exchanged directly.

3.1 Single BM scheme (SBMS)

This is the simplest and most direct way to reduce the BM size. Because the compression process is base on each single BM, we name it single BM scheme (SBMS).

The primary purpose of BM exchange is to tell other peers how one peer's buffer is filled. As mentioned above, a bitmap of a {0,1} sequence is introduced to represent the filling statue and is exchanged between peers. In a practical streaming system, chunks at positions near the buffer head have been in the system much earlier than those near the buffer tail; hence the position closer to the head has larger chance to be filled. In other words, a bit position at buffer head has larger probability to take the value 1 while a bit position at buffer tail has larger probability to be 0. Figure 2 is measured filling probability vs position of UUSee.

SBMS just takes advantage of this fact of bitmap to compress each single BM, and especially, both compression and decompression process don't depend on last or any other BM.

The BM compression methods used in engineering practice generally belong to this category. In midterm of 2009, we firstly find certain compression methods are applied in UUSee and PPLive. In later of 2009, we success in analyzing the compression process of UUSee, where an original bitmap of more than 400 bits can be reduce to 17.5 bytes long on average by certain variant of LZ and run-length algorithms and each BM can be decompressed independently. The data used in our following theoretic analysis and simulation is based on the decompression of the measured BM of UUSee. In an investigation shortly afterwards, we get to know PPLive adopt 2 level huffman algorithm to process their BM. In brief, both of them are base on different positions' 0/1 probability statistics in the original bitmap, while the strong correlation between adjacent BMs are not realized.

In general, SBMS merely use the general data compression algorithms in compressing the BM and doesn't make any other breakthrough and innovation. Of course the biggest strength lies in its simpleness.

3.2 Single Peer's BM scheme (SPBMS)

We name it as the single peer's BM scheme (SPBMS) as the compression is based on the correlation between two continuous sending BMs of a single peer.

1) The Principle

A seemly trivial observation in buffer filling is that, once a buffer position is filled, it will be filled forever. In other words, only those bit positions with value 0 in current reported bitmap may change their values to 1 at following reported bitmaps. Thus this seemly trivial observation will introduce a new non-trivial compression philosophy: Single Peer's BM scheme (SPBMS). In this scheme, a peer will stop reporting those positions that have ever been notified with value 1 at previous BM sending.

Thus, the SPBMS follows this principle:

Principle 1: A peer needs not to send message about a position further once a value 1 has been sent in this position.

Then we will have different choices on how to send the remaining positions with value 0 in previous bitmap. It seemingly has less amount of information if we just send the positions with value variation from 0 to 1 in current BM. However, it should be noted that there are two types of information to be sent for each buffer position: the value 0 or 1 and the location. For instance, bitmap itself just uses the offset and bit sequence to make the value and position self-explanatory. If only the positions with value variation are to be sent, we have to code the locations information and it seems inefficiency and complex. On the other hand, in SPBMS, the current bitmap are condensed into a short one which consists of bit value sequence of all the positions with value 0 in previous bitmap.

In order to locate the position of each bit, a special integer set named as support set (SS) is introduced. In theory, the SS is a locations (*chunk id*) set of all the chunks which have not buffered in the BM sender. All location values in the SS are in ascending order. In each round, sender will extract the bit value sequence as the compression sequence from its current bitmap according to the locations starting from head of SS, then all the locations which have buffered in current bitmap will be removed from the SS; the receiver will retrieve the bitmap from the compressed bit sequence according to the locations beginning from the head of SS sequentially, then all the locations with bit value 1 in current received bitmap will be removed from the SS.

For correct encoding and decoding, the SSs will be updated continuously based on each bitmap, and both the sender and receiver must keep the same support set at the same time. This means a decompression peer has to keep communication with the compression peer from the very beginning, and in practice such a requirement can be met by either periodically announcing a compressed BM or the current support set.

The proposal proposed by [2] in 2009 is one typical implementation of SPBMS. Simulation proofs it can get more advantage than the implementations used by PPLive or UUSee. To the best of our knowledge, they for the first time put forward the compression method specific to BM exchanging. However, their work isn't thorough enough. They give neither a strict analysis in theory nor a proof in practice on their algorithm. Most of important, because in SPBMS what a peer sends always beyond what the receiving peer needs in BM exchanging, any solution based on it is not an ultimate one.

Therefore, a question of the optimum scheme is present. What are the best compression scheme and solution; and how many gains can be brought with it over other solutions? All of these questions are very interesting and worth deeply studying. They are not only the key points that we will make effort to resolve, but also the innovation of this paper.

3.3 Paired peers' BM scheme

In this paper, we for the first time put forward a BM compression scheme called paired peer's BM scheme (PPBMS). The name comes from the fact it can condense a bitmap by removing the redundant information between the pairwise peers' exchanged BMs besides eliminating the redundancy between a peer's adjacent BMs.

1) The conception and principle

On closer inspection of interactive behavior of paired peers, we find another seemingly insignificant fact: once a buffer position is filled, its peer will not care for this position any more whether the same position is filled or not in other peers. In other words, a peer only concerns about the situation of those buffer positions with value 0 in his own bitmap in other peers. Thus this observation along with the principle 1 will introduce another innovative compression philosophy: Paired Peer's BM scheme (PPBMS). In this scheme, a peer will stop reporting those positions that have ever been notified with value 1 in either his own previous sending BM or his connected peer's BM. Different from SPBMS, PPBMS takes a peer's counterpart's needs into consideration and the result compressed BM is specific to the counterpart of a sender. Intuitionally the scheme may remove the redundancy to a large extent.

Thus, the PPBMS follows another principle besides principle 1:

Principle 2: A peer needs not to send message about the position to his paired peer further after receiving a value 1 in this position from his paired peer.

Due to the same reason discussed above in SPBMS, in PPBMS the peer will send all the bit value sequence extracted from his bitmap ready to be sent corresponding to all the positions with value 0 in either a sender's previous bitmap or the latest received bitmap. The SS with ascending sorted values is also necessary for correct compression and decompression, but has some fundamental different meaning and management mechanism. First, the SS in PPBMS is a locations (*chunk id*) set of all the chunks which have not buffered in both paired peers; secondly, both roles (BM sender/receiver) of a peer will share the same SS; thirdly, considering the SS synchronization, the processes of BM exchange in two directions are dependent to each other. Except those differences, SS uses the similar dynamic formation/update mechanism.

For correct encoding and decoding, the SS will be updated continuously based on each bitmap exchange in both directions, and both the peers must keep the same support set at the same time. This requires the BM exchange takes place in single process from the very beginning in both directions between paired peers. In engineering design, considering network conditions (loss and delay) and concurrency of BM exchange, we can decouple the correlation by adopting multiple independent SSs, each of which is for one time of BM exchange. The core idea is: based on both the receiver's latest confirmation about which BM (called local BM reference or LBMR) sender has reported and the latest received BM from the receiver (called counterpart BM reference or CBMR), the sender compresses his new BM and sends it out. Of course, along with the compressed sequence, other information such as unique indexes of LBMR and CBMR needs to be transferred for locating the SS decompression needs.

2) The Basic Protocol Progress

The basic protocol is the same as that of SPBMS except that in PPBMS both roles (BM sender/receiver) of a peer share the same SS, hence the details of progress is omitted here.

3) The consistency in the SS of paired peers

Under the same assumptions as in SPBMS except that in PPBMS both roles (BM sender/receiver) of a peer share the same SS, we define $|v|=k$ for a binary sequence $v=(v_1,...,v_k)$. Initially, SS $L_A=L_B=$\{all location\}. Assuming BM exchange is started at time $t$ as peer A sends his BM to B, then

- A sends $\varphi_A(t)$ and bit sequence $(b_0,...,b_{N-1})$ to B
- $L_A=\{l \geq \varphi_A(t)$ & the locations of the chunk that peer A has not buffered at time $t\}$
- $L_B=\{l \geq \varphi_A(t)$ & the locations of the chunk that peer A has not buffered at time $t\}$
- So $L_A=L_B= \{l \geq \varphi_A(t)$ & the locations of the chunk that peer A has not buffered at time $t\}$

Assume peer A sends to B at time $t_0$ and $L_A(t^+_0)=L_B(t^+_0)$, then B sends to A at time $t$, then

- B sends $\varphi_B(t)$ and bit sequence $v=(v_1,..., v_m)$ to A
- Let $L_B=\{l \in L_B(t^+_0) : l \geq \varphi_B(t)\}$ in ascending ordered, $L_B(t^+)=\{l_j \in L_B: v_j=0,$ if $j<|v|\}$
- Let $L_A=\{l \in L_A(t^+_0) : l \geq \varphi_B(t)\}$ in ascending ordered, $L_A(t^+)= \{l_j \in L_A: v_j=0,$ if $j<|v|\}$
- So we still have $L_A(t^+)=L_B(t^+)$

We can draw the same conclusion if B sends to A at time $t_0$ and $L_A(t^+_0)=L_B(t^+_0)$, then A sends to B at time $t$.

Therefore, the consistency in the SS is proved.

IV The Analysis of Different Schemes in Theory

In this section, we will discuss the compression limit ratio of different schemes, as well as the protocol overhead of BM exchange in theory. All the issues are very interesting and important not only to us but alsoto the system's designer and developer.

4.1 Theoretic Analysis

In our previous research [3], we find and validate buffer filling in P2P streaming system is a stationary process following certain s curve. Fig.1 depicts the S curve of UUSee client measured in April 2009. The whole content buffer is broken down into many sequential blocks (chunks) as the horizontal axis shown, and the vertical axis is the filling probability of each block position. For a given stable peer, the s curve is independent of any observation time and the offset lag to any other peer. With this knowledge, the compression rate of different schemes can be deduced based on information theory.

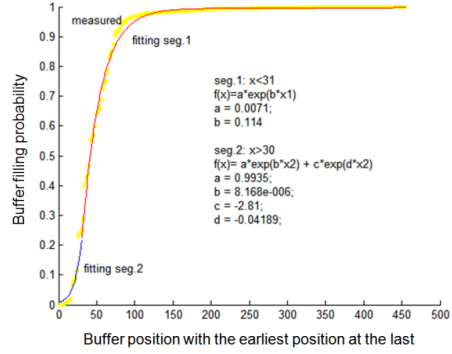

Figure1. the diffusion S curve of UUSee

For further theoretic analysis, we introduce the following definitions and assumptions (see Fig.4) :
a) BM includes an *offset φ* and a *bitmap ($b_0,...,b_{N-1}$)* with the buffer length is *N chunks*;
b) The buffer filling probability is {$p_i$, $0 \leq i \leq N-1$}
c) *T* is the period of BM exchange of a peer, and *τ* is the exchange delay between adjacent peers of both peers' from one peer, say B, to another: if say peer B sends its BM to its counterpart A at time $t+iT$, i=0,1,..., then A sends to B at time $t+iT+\tau$, $0<\tau\leq T$, i=0,1,...;
d) To simplify the analysis, we assume the both A and B have the same playback delay, i.e. offset $\varphi_A = \varphi_B$ at any time.

1)         The analysis of SBMS

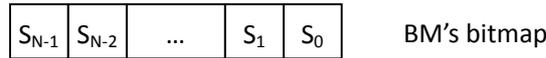

Figure 2. Information model of SBMS

According to theory of information, we can regard the bitmap of a BM as the combination of N binary source { $S_i$, $0 \leq i \leq N-1$}. As each chunk's downloading can be affected by many factors including network conditions, local data sharing, chunk fetching policy, etc., for peer in stable condition, it is reasonable to assume the binary sources are independent to each other.

Therefore, we have the information quantity of compressed bit sequence in SBMS

$$H_{SBMS} = \sum_{i=0}^{N-1} H(S_i)$$

2)         The analysis of SPBMS

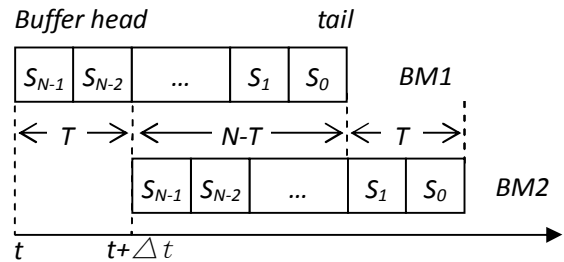

Figure 3.   Information model of SPBMS

Reference to the two continuous BMs of a peer in Figure

3, any two binary sources with T chunks interval between $BM_1$ and $BM_2$, ($S_i$ in $BM_1$, $S_{i+T}$ in $BM_2$) $T \leq i \leq N-1$ have strong correlation. Because the $S_i$ and $S_{i+T}$ correspond to the same chunk, if $S_i=1$ in $BM_1$, then $S_{i+T}$ in $BM_2$ must be 1; if $S_i=0$ in $BM_1$, then $S_{i+T}$ in $BM_2$ may be 1 with certain probability. For simplicity, we define, *p(i=1)* as the probability the chunk at position *i* has been buffered in BM1, $p(j=1)$, where $j=i+T$, as the probability the chunk at position $j$ has been buffered in BM2, and symbol $p(j=1/i=0)$ as the condition probability that the same chunk at position $i$ in $BM_1$ is not buffered but buffered at position $j$ in $BM_2$. Considering the ergodicity of buffer filling, we have following equation:

$$p(i=1) + p(i=0)p(j=1/i=0) = p(j=1)$$

Let $p_i=p(i=1)$, $q_{i,j}=p(j=1/i=0)$, above equation can be written as:

$$p_i + \overline{p}_i q_{i,j} = p_j$$

Then, we get $q_{i,j} = \dfrac{p_j - p_i}{\overline{p}_i}$

Therefore, the information quantity of the compressed bitmap in SPBMS is

$$H_{SPBMS} = \sum_{i=0}^{T-1} H(S_i) + \sum_{i=0}^{N-T-1} H(S_{i+T}|S_i) = \sum_{i=0}^{T-1} H(S_i) + \sum_{i=0}^{N-T-1} p(i=0) H(q_{i,j})$$

3) The analysis of PPBMS

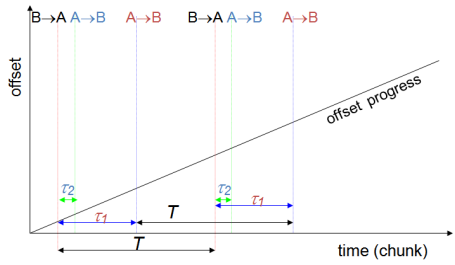

Figure 4. the BM exchange $\tau$ between peers in PPBMS

It is a little complex to analyze the PPBMS. Let's recall that in PPBMS only the statuses of those positions which have not been buffered in both paired peers' bitmaps should be reported. Seeing Fig.4, first, let's think about peer A sends his BM to peer B at time $t$. A's previous BM (LBMR) occurs at time $t-T$ in all cases, while occurring time $t-\tau$ of B's last BM (CBMR) will be different with different $\tau$ selection. So, the probability to exchange information about a position k in A's BM is *Pr(position k-T is 0 in LBMR)×Pr(position k-$\tau$ is 0 in CBMR)*. On the other hand, at the time $t$ peer B sends his BM to peer A, B's last BM occurs (LBMR) at time $t-T$ in all cases, while A's last BM (CBMR) time $t-(T-\tau)$ will be different with different $\tau$ selection. So, the probability to exchange information about position k in B's BM is *Pr(position k-T is 0 in the LBMR) Pr(position k-(T-$\tau$) is 0 in CBMR)*. Obviously, in PPBMS, the entropies in two directions between paried peers will be different, and related to the exchange time $\tau$.

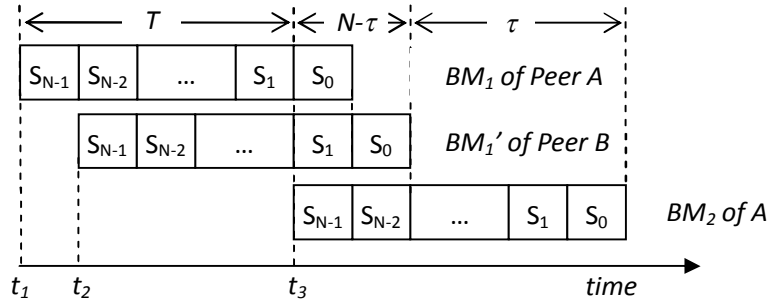

Figure 5. Information model of PPBMS

See the information model of PPBMS in Fig.5, peer A will send his BM to B $\tau$ time later after receiving peer B's BM, and correspondingly peer B will send his BM to A $T-\tau$ time later after receiving peer A's BM.

In the case of A to B, $BM_1$ and $BM_1'$ are the LBMR and CBMR of peer A respectively, and $BM_2$

are ready to be sent. Clearly, the information quantity consists of three parts: the complete appending part which doesn't exist in LBMR and CBMR; the partially appending part which is ever reported in CMBR but not in LBMR; the updated part which are ever reported in both LBMR and CBMR. According to theory of information, we can deduce the amount of information

In direction of A to B:

$$H_{AB} = \sum_{i=0}^{\tau-1} H(S_i) + \sum_{i=\tau}^{T-1} p(i-\tau=0)H(p_i) + \sum_{i=T}^{N-1} p(i-\tau=0)p(i-T=0)H(p(i|i-T=0))$$

Correspondingly, in the reverse direction B to A

$$H_{BA} = \sum_{i=0}^{(T-\tau)-1} H(S_i) + \sum_{i=T-\tau}^{T-1} p(i-(T-\tau)=0)H(p_i)$$

$$+ \sum_{i=T}^{N-1} p(i-(T-\tau)=0)p(i-T=0)H(p(i|i-T=0))$$

In fact, we are more interested in the average entropy between paired peers, i.e.,

$$H_{PPBMS} = (H_{AB} + H_{BA})/2$$

4.2 the Information Quantity based on UUSee's Diffusion S curve

In general, the diffusion (S) curve is easy to obtain by P2P network measurement and S curve fitting, while we can't get the closed-form solution for the information quantity of different schemes. Instead, we adopt numerical analysis method in our further study.

Based on our BM trace, we fit the diffusion S curve as the 2-segment filling curve shown in Fig.1. Substituting the fitting curve into the information quantity functions, we can easily calculate the information content.

1) Curve of PPBMS

In general, the parameters involved in above function, including the buffer width N(=456), cycle of BM sending and buffer filling probability, are all P2P system design parameters, while the exchange time delay $\tau$ used in PPBMS is not. So, before comparing different schemes, we'd like to discuss one interesting question, i.e. how the value of exchange time delay $\tau$ will affect the information quantity of PPBMS and what is the minimum value.

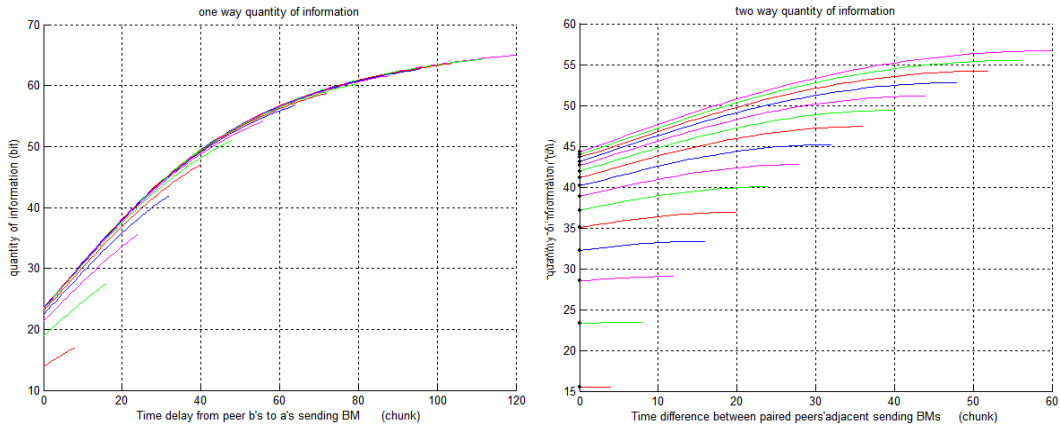

(a) The one-way curve     (b) two-way on average curves

Figure 6. the information content of PPBMS

Fig.6 depicts the information content curves of PPBMS in one-way and two-way average mode for different $\tau$ and $T$. In this figure, the horizontal axis is the exchange time delay $\tau$, and the vertical axis is the information quantity; different line stands for different BM sending cycle $T$, which is successively rising with 8 chunks for curves from bottom to top respectively; the offset lag = 0 between both peers in all cases. It needs to point out that the information content of the paired-peers are symmetric with the center $\tau=T/2$ in the Fig.6(a). Dividing the sum of the both symmetric values by 2, we get the curves in Fig.6(b). Observing these curves, we have following findings: *i)* The information quantity is a certain increment function of BM's sending cycle *T* for any given exchange time delay *τ*; *ii)* The information quantity is also a certain monotone increment function of *τ* for any given BM's *T*; *iii)* For any given period *T*, the information content has the minimal value, which will be quoted in following comparison, when exchange time delay *τ=0*; *iv)* The shapes of both one-way and two-way curves seem alike except the two-way curve is more flat.

We think the shape of the curves is mainly affected by the diffusion S curve, however, the more deep reasons are to be left to our future research.

2) Comparison among different BM schemes

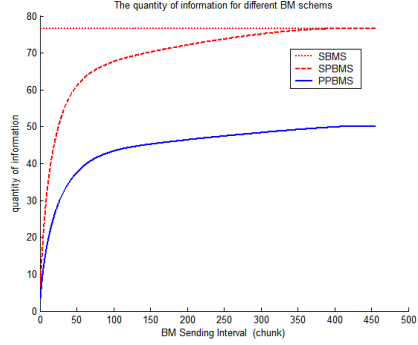

Figure7. Comparison among all schemes with UUSee

We draw the information content curves of SBMS, SPBMS and PPBMS in one Fig.7 for performance comparison. Obviously, SBMS has a constant value of 77 bits; the curves of both SPBMS and PPBMS seem to be certain logarithmic functions of BM sending period *T*. Significantly, for any given BM sending cycle *T*, our PPBMS has absolutely advantage over other compression schemes. For accurate comparison, we calculate the values with the most often used samples of T in table 1, where we can see, from the compression rate perspective, our PPBMS can achieve the best compression ratio and it has a gain of 42% over SPBMS, and 68% over SBMS on average in theory. Because PPBMS nearly remove all redundant information of BM, the values with *τ=0* can seem as the compression limit in theory for P2P streaming system.

Table 1. the theoretic information quantities and the gain of PPBMS

|  | BM sending period (chunks) | | | | |
| --- | --- | --- | --- | --- | --- |
|  | T=8 | T=16 | T=24 | T=32 | Average |
| SBMS(bit) | 77 | 77 | 77 | 77 |  |
| SPBMS(bit) | 28 | 41 | 49 | 54 |  |
| PPBMS(bit) | 15 | 23 | 29 | 32 |  |
| gain over SPBMS | 0.45 | 0.43 | 0.42 | 0.41 | 0.42 |
| gain over SBMS | 0.80 | 0.70 | 0.63 | 0.58 | 0.68 |

3 BM protocol overhead comparisons

We define BM protocol overhead as the signal overhead introduced by BM exchange in term of network traffic in a certain period. As we know, with small BM sending period *T*, we have small compressed BM size but high frequency of BM exchange; while from large *T*, low frequency of BM exchange but big compressed BM size may be resulted. So, we want to investigate the overhead and get the tradeoff between the BM sending cycle T and the compression BM size.

All three schemes' overheads, which are calculated with the information quantity divided by its corresponding BM sending period T, are dawn in Fig.8, and table 2 shows the overheads with typical BM sending cycle T. Without any knee points expected, all the overheads' curves are monotone descent function of BM sending cycle *T*.

Comparing all the overhead curves, although our PPBMS overhead reduction seems mainly in the head, in fact even at T=400 the reduction gain over SPBMS still reach up to 35%.

According to libo [1], increasing the frequency of BM exchange is helpful to improve the P2P system performance. Setting the overhead of SBMS at T=22 as the criterion, we can see the BM sending frequency can be improved 2 times faster in SPBMS while nearly 20 times faster in our PPBMS. Therefore, the significant compression advantage of PPBMS has been proved.

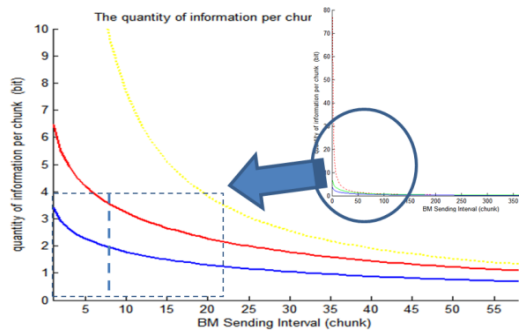

Figure8. Comparison among the overhead of all schemes

Table 2. the theoretic overhead and the gain of PPBMS
signal cost (bit/per chunk time)

| schemes | BM Sending Cycle | | | | | average |
|---|---|---|---|---|---|---|
| | T=8 | T=16 | T=24 | T=32 | T=40 | |
| SBMS | 9.59 | 4.79 | 3.20 | 2.40 | 1.92 | |
| SPBMS | 3.53 | 2.57 | 2.05 | 1.70 | 1.45 | |
| PPBMS | 1.94 | 1.46 | 1.19 | 1.01 | 0.88 | |
| gain over SBMS | 0.45 | 0.43 | 0.42 | 0.41 | 0.40 | 0.42 |
| gain over SPBMS | 0.80 | 0.70 | 0.63 | 0.58 | 0.54 | 0.63 |

V Simulation and validation with measurement BM trace

Based on our actual measured UUSee BM interactive trace, we simulate different operational BM schemes' implementations and try to use the quantitative results to validate the validity of our PPBMS.

The BM trace used in our simulation is extracted from the session with the longest life in our UUSee client measurement dataset in April 2009. The trace with 4689 BM records lasts for about 1 hour and 16 minutes, and all records are compressed by UUSee with certain LZ and run length algorithm according to the SBMS. By reverse engineering method, we are success to decompress it and get the non-compressed BM sequence, which is used in the following simulations of SPBMS and PPBMS at last. According to our reverse engineering analysis, on average, the length of the non-compressed bitmap of a BM is about N=54.4 bytes which is compressed to 17.5 bytes long as Fig.10(a) shown, the BM sending cycle is about T=20 chunks, and the exchange time $\tau$=5chunks.

1 the SPBMS implementation

The method proposed by [2] in 2010 is one typical implementation of SPBMS. Seeing Fig.9, according to this method, only following information needs to be reported in the compressed BM: *i)* the current statuses of chunks which are not buffered in previous BM; *ii)* the statues of the appending chunks in current BM. Furthermore, we try to compress the bit sequence output by SPBMS with run-length/Huffman and arithmetic coding, and expect a short bit sequence. It needs to point out we have removed all the duplicate BM records before feeding the trace into the SPBMS implementation.

The simulation result is shown in Fig.10(a) and table 3. As we can see, the average BM length with SPBMS is 4.03 bytes; with this output sequence, the length with run-length, Huffman and arithmetic coding is 9.2, 3.8 and 5.7 bytes. It seems a contradiction that the simulation result has smaller size than the theoretic limit, but it should be noted that the simulation result in the table is only the bitmap information of a BM. Considering the complexity of real network conditions including packet loss and retransmission, network delay, as well as the BM's timeliness, some other bytes, including the BM's offset, its reference BM and the confirmed received BM, should be introduced to locate the bitmap of a BM. In general, these location information needs about more than five bytes. In addition, there is nearly no room to further compression with RL,Huffman or AC algorithms after SPBMS according to the simulation. Reference to Fig.12(a), this maybe mainly ascribed to fact that most symbols of the SPBMS sequence are random enough.

Table 3. The size statistics in simulation
unit: bytes

| Schemes | The Limit | Basic length | Run-Length (RL) | Huffman after RL | Arithmetic Coding |
|---|---|---|---|---|---|
| Original | - | 54.4 | - | - | - |
| SBMS | 9.59 | 17.5 | - | - | - |
| SPBMS | 5.70 | 4.03 | 9.3 | 3.8 | 5.7 |
| PPBMS | 3.27 | 2.6 | 6.4 | 2.6 | - |

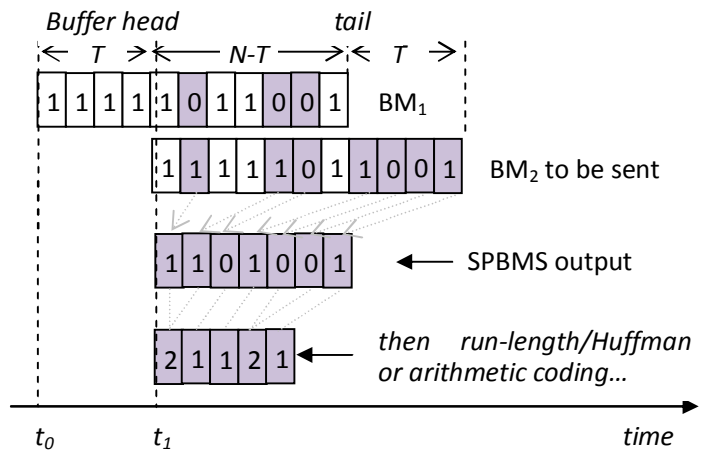

Figure9. the scenario of SPBMS

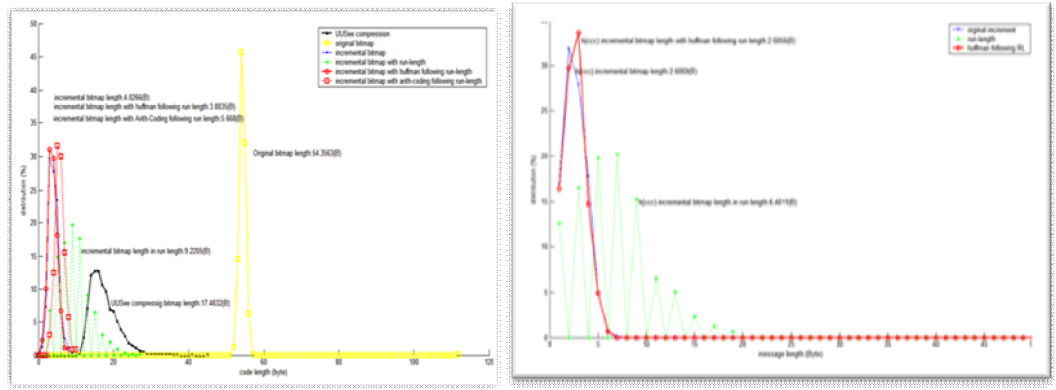

(a) SPBMS simulation  (b) PPBMS simulation

Figure10. the simulation results

2 the PPBMS implementation

According to PPBMS, in a typical implementation, the sender only needs to report the statuses of those positions which don't buffer in both LBMR and CBMR. For better understand the process, we draw a scenario in Fig.11.

In our simulation, for simplicity, we assume there is no transmission delay and loss in BM exchange, thus we can simply use the last sent BM and last received BM as the LBMR and CBMR respectively.

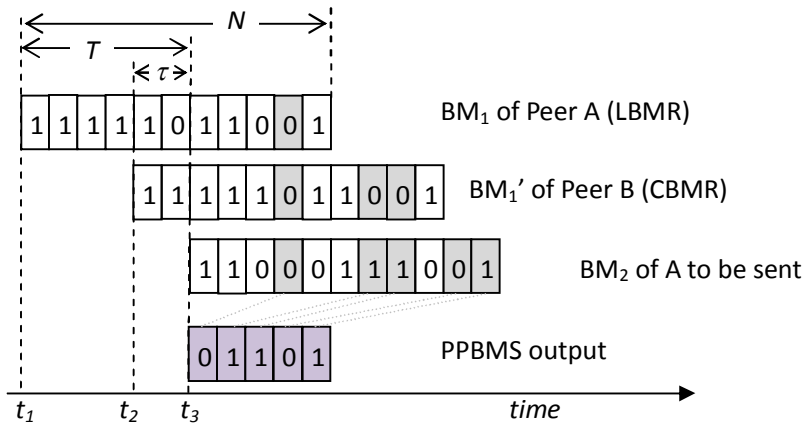

Figure 11. The scenario of PPBMS

The simulation result is shown in Fig.10(b) and table 3. The average BM length with PPBMS is 2.6 bytes; with this output sequence, the length with run-length and Huffman is 6.4 and 2.6 bytes respectively. Due to the same reason as discussed in above section, the simulation result has smaller size than the theoretic limit. In real implementation, some extra location information including the BM's offset, its reference BM and the confirmed received BM, which usually needs about more than five bytes, should be appended. In addition, no margin is left for RL and Huffman algorithms according to the simulation, and Fig.12(b) shows the symbols distribution.

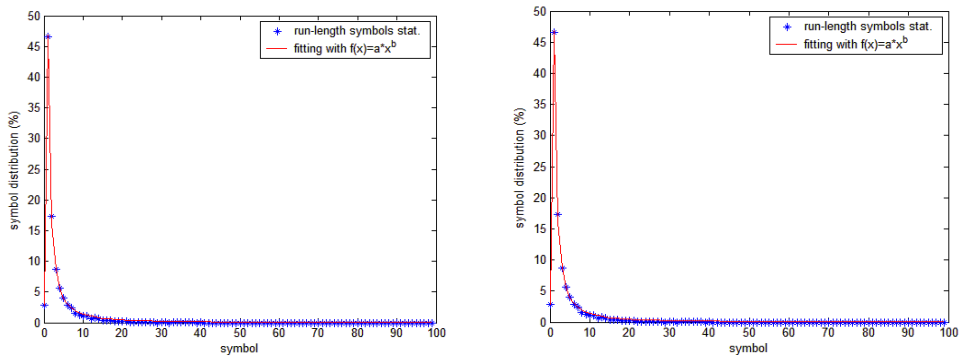

(a)  Symbol distribution in SPBMS      (b) Symbol distribution in PPBMS

Figure12. The symbol distribution

Through our simulation, we can see our PPBMS get more than 35% compression advantage over SPBMS in the same conditions and the quantitative result validates the validity of our PPBMS.s

VI Summary

BM compression is an interesting and important issue which has never been thought about and discussed seriously and thoroughly. In this paper, we discuss different kinds of compression schemes and propose a more powerful and engineering operational compression scheme – PPBMS; Through profound analysis in theory of information content and simulation with our measured BM trace of UUSee, we validate the validity and superiority of our PPBMS in compression ratio. In our further study, we will continue to resolve other issues presented in this paper.

Reference
[1]  Chen Feng, Baochun Li, Bo Li. "Understanding the Performance Gap between Pull-based Mesh Streaming Protocols and Fundamental Limits," in the Proceedings of IEEE INFOCOM 2009, Rio de Janeiro, Brazil, April 19-25, 2009.

[2] Guangqing Deng, Yunfei Zhang, Chunxi Li, Changjia Chen, "A bitmap coding method for P2P streaming protocols", in Proc. of CAR, 2010

[3] Y. Chen, C. Chen and C. Li, "Measure and Model P2P Streaming System by Buffer Bitmap", HPCC'08, Sept. 2008,

[4] D. Qiu and R. Srikant, Modeling and performance analysis of bittorrent-like peer-to-peer networks. In: Proceedings of SIGCOMM 2004, Portland, Oregon, USA, 30 Auguest-3 September, pp. 367-378

[5] F. Piccolo, and G. Neglia, The effect of heterogeneous link capacities in bittorrent-like file sharing systems. In: Proceedings of Hot-P2P 2004, Volendam, The Nederlands, 8 October 2004, pp. 40-47

[6] F. Clevenot, P. Nain, and K. Ross, Multiclass p2p networks: Static resource allocation for service differentiation and bandwidth diversity, Performance Evaluation 62 (1-4) (2005) 32-49

[7] R. Kumar, Y. Liu, and K. Ross. Stochastic fluid theory for P2P streaming systems. In: Proceedings of IEEE INFOCOM 2007, Anchorage, AK, 6-12 May 2007, pp. 919-927

[8] S. Liu, R. Zhang-Shen, W. Jiang, J. Rexford, and M. Chiang, Performance bounds for peer-assisted live streaming. In: Proceedings of Sigmetrics 2008, Annapolis, MD, USA, 2-6 June 2008, pp. 313-324

[9] Pai V, Kumar K, Tamilmani K, et al. Chainsaw: eliminating trees from overlay multicast. Proceedings of IPTPS 2005, Peer-to-Peer Systems IV. Vol. Volume 3640/2005. Springer Berlin / Heidelberg, 2005:127-140.

[10] Zhang X, Liu J, Li B, et al. Coolstreaming/donet: a data-driven overlay network for peer-to-peer live media streaming. Proceedings of IEEE/INFOCOM'05, Miami, USA, 2005:2102-2111

[11] Magharei N, Rejaie R, Guo Y, Mesh or Multiple-Tree: A comparative study of live P2P streaming approaches. Proceedings of Proceedings of IEEE INFOCOM'07, 6-12 May 2007, Anchorage, AK, 2007:1424-1432

[12] D. Ciullo, M. A. Garcia, A. Horvath, E. Leonardi, M. Mellia, et al. Network awareness of p2p live streaming applications. in Proceedings of IPDPS 2009, Rome, Italy, 25-29 May 2009, pp. 1-7

[13] Hei X, Liang C, Liang J, et al. A measurement study of a large scale P2P IPTV system, IEEE Transactions on Multimedia, 2007, 9(8):1672-1687